%Paper: astro-ph/9406062
%From: BELLAZZINI@ASTBO3.BO.ASTRO.IT
%Date: Wed, 22 JUN 94 15:17 N

% the two table are appended at the end of the paper, extract them and
%"tex" them separately. At the very end the macro BASINEW.TEX is appended
%(for tables only)

\magnification=1200
\hoffset=0.0 true cm
\voffset=0.0 true cm
\vsize=21.5 true cm
\hsize=16.3 true cm

\baselineskip=15pt     %%%%  change these to double spacing before submitting
\parskip=5pt           %%%%  it; also put page ejects where needed

\parindent=22pt
\raggedbottom

\def\pp{\noindent\parshape 2 0.0 truecm 17.0 truecm 0.5 truecm 16.5 truecm}

\font\bigbf=cmb10 scaled \magstep2

\def\etal{{\it et~al.~}}
\def\lsim{\hbox{ \rlap{\raise 0.425ex\hbox{$<$}}\lower 0.65ex\hbox{$\sim$} }}
\def\gsim{\hbox{ \rlap{\raise 0.425ex\hbox{$>$}}\lower 0.65ex\hbox{$\sim$} }}

%%%%%%%%%%%%%%%%%%%%%%%%%%%%%%%%%%%%%%%%%%%%%%%%%%%%%%%%%%%%%%%%%%%%%%%%%%%%%%

\null\vskip 1.0 truecm

\centerline{\bigbf BLUE STRAGGLERS IN GALACTIC GLOBULAR}
\bigskip
\centerline{\bigbf CLUSTERS: PLAYING WITH SPECIFIC QUANTITIES}
\bigskip
\bigskip

\centerline{\bf F.R. Ferraro$^1$,$$ F. Fusi Pecci$^1$ $$ \& M. Bellazzini$^2$}

\bigskip
\bigskip

\centerline{\it $^1$ Osservatorio Astronomico di Bologna, Via Zamboni 33,
I-40126 Bologna, Italy.}
\centerline{\it $^2$ Dipartimento di Astronomia, Universit\'a di Bologna,}
\centerline{\it Via Zamboni 33, I-40126 Bologna, Italy.}

\bigskip
\bigskip
\bigskip
\centerline{e-mail: 37907::ferraro}
\centerline{e-mail: ferraro@alma02.astro.bo.it}
\centerline{fax: 39-51-259407}

\bigskip
\bigskip

\centerline{Submitted to {\it Astronomy and Astrophysics}}
\bigskip

\centerline{Accepted: June, 6, 1994}

\bigskip
\vfill\eject
\centerline{ABSTRACT}
\bigskip
\noindent {
We present preliminary results obtained from the comparison of the
specific frequencies of Blue Straggler Stars (BSS) detected so far
in a sample of 26 Galactic globular clusters.

The number of BSS seems to increase almost linearly with increasing the
amount of sampled light in loose clusters, while it drops abruptly for
clusters having intermediate-high central densities.
In particular, a simple interpretative scenario where the
BSS in loose clusters are produced from primordial binaries and
those detected in high density globulars are due to star
interactions leading to binary formation, merging. etc.
seems compatible with these early results.

The possibility that this observational evidence could be
ascribed to systematic biases mostly related to the increasing difficulty
to detect BSS candidates with increasing cluster concentration, is also
discussed.}

\bigskip\noindent
{\it Thesaurus: 03.05.1; 19.20.1}

\bigskip

\centerline {1. INTRODUCTION}
\bigskip\noindent
Many new results are now supporting the claim that dynamical evolution
of globular clusters (GCs) can affect the evolution of their stellar
populations.
In fact, both the integrated cluster colors
(Djorgovski \etal 1988, Piotto \etal 1988, Bailyn \etal 1989, 1992,
Auri\`ere \etal 1990, Stetson 1991, Cederbloom \etal 1992, etc.)
and the properties of special individual stars
(Horizontal Branch (HB): Buonanno \etal 1985, 1986, 1991, Ferraro
\etal 1992a, Fusi Pecci \etal 1993a,
Blue Stragglers (BSS): Fusi Pecci \etal 1992, Sarajedini 1992,
Ferraro \etal 1993),
confirm the existence of dynamically induced variations in the evolution
of many cluster members.

Moreover, as recently reviewed by Bailyn (1993), there is now a variety
of observational results which are direct or indirect evidences of the
existence and impact of various kinds of binary systems in globular
clusters. Physical interactions between stars are quite probable,
and significant modifications of the populations of both single and
binary stars can naturally be expected.

As a consequence, it seems natural to conclude that
a combination of several phenomena somehow related to
binaries and/or environment may be responsible for the stellar population
gradients, for the properties of the BSS and HB stars, for the existence of
rare or even ``exotic'' objects, for the production of millisecond pulsars
and X-ray sources, etc.

In particular, as widely reviewed by many authors in several recent papers
(Nemec 1989, 1991, Leonard 1989, Fusi Pecci \etal 1992, Bailyn 1992, Stryker
1993 and references therein), there is a growing belief that binaries via
various mechanisms (e.g. coalescence, merging, interaction, capture) can be
related to the origin of the BSS detected in Galactic GCs.

The recent growing body of observations reporting new detection of BSS
candidates in Galactic GCs from the ground (see Fusi Pecci \etal 1992, 1993b
for references) and with HST just at the center of highly concentrated
globulars (Auri\`ere \etal 1990, Paresce \etal 1991, De Marchi \etal 1993,
Ferraro and Paresce 1993, Lauzeral \etal 1993,
Bailyn and Mader 1993) has renewed the interest of the astronomical community
on this topic, strongly emphasizing the need for appropriate inter-cluster
comparisons in order to detect any possible correlation between the cluster
overall parameters and the observed BSS properties.

In this respect, we have already presented (Fusi Pecci \etal 1992,
Ferraro \etal 1993) some observational evidences supporting the existence
of possible correlations between cluster structural parameters and BSS
properties. In particular, we found that:
\par\noindent
\item{\it (i)} The BSS Luminosity Function (LF) for low density GCs
($Log\rho_0<3$) turns out to be different from that obtained for the BSS
detected so far in highly concentrated GCs ($Log\rho_0>3$), at $3\sigma$
level.
\par\noindent
\item{\it (ii)} There are some (weak) indications that the ridge line of the
BSS sequence in the CMD is progressively red-shifted compared to the
bright extension of the Zero Age Main Sequence with increasing
metallicity.
\par\noindent
\item{\it (iii)} In M3, the BSS radial distribution normalized to the
sampled light presents a strongly different behaviour with respect to that of
``normal'' stars in the same magnitude range, and there
is a lack of BSS in an intermediate annular region of the cluster
which could be ascribed to the existence of two
different populations of BSS within the same cluster or to segregating
effects in the BSS production and/or survival (Ferraro \etal 1993).

\smallskip\noindent
In this paper, we report on the first quantitative comparison  of the
rate of production/survival of BSS in the Galactic GCs
where such objects have actually been detected so far. Since it is
highly probable that the available sample of BSS (625 listed in Table 2
of Fusi Pecci \etal 1993b) is still substantially incomplete, especially for
the dense clusters, important caveats on the use of BSS specific quantities
are also discussed.

\bigskip\noindent
\centerline {2. THE DATA SET}
\bigskip\noindent
Table 1 presents the list of 26 GCs containing BSS for which
$r_c$ -- ~the core radius~ -- and $r_t$ -- ~the tidal radius~ -- are known,
and precise information on the area covered by the observations could be found.
The selection of the BSS have been made by visual inspection of the
published Color Magnitude Diagrams, whose reference has been reported in
column 9 (see also Fusi Pecci \etal 1992). In particular, for each GC we list:
\par\noindent
(1) The cluster name.
\par\noindent
(2) The logaritm of the central mass density ($Log\rho_0$) in
$M_{\odot} pc^{-3}$ from Webbink (1985).
\par\noindent
(3) The concentration parameter ($c=Log~r_t/r_c$). To keep homogeneity
in the data-source, this value has been computed starting from the observed
values of $r_t$ and $r_c$ listed in the Webbink compilation. Note that the
use of any other list of similar data would leave the present results
and discussion substantially unchanged.
\par\noindent
(4) The absolute magnitude ($M_V$) computed scaling the apparent integrated
magnitudes listed by Webbink (1985) for the appropriate GC distance modulus
(see below). Small differences between the values listed here and those
adopted in our previous papers (Fusi Pecci \etal 1992,1993b) are actually
irrelevant for the present purposes.
\par\noindent
(5) The metal abundance ($[Fe/H]$) taken from Zinn (1985).
\par\noindent
(6) The bolometric luminosity of the surveyed area in unity of $10^4
L_{\odot}$ (hereafter $L_S$) defined  below.
\par\noindent
(7) The number of the adopted BSS ($N_{BSS}$) selected on photometric
grounds (see Fusi Pecci \etal 1992).
\par\noindent
(8) The {\it specific frequency} ~~$S4_{BSS}$, see Sect. 3.
\par\noindent
\par\noindent
(9) The {\it crowding parameter} ~~~$F$, defined and discussed in Sect. 3.2
to estimate somehow the possible influence of crowding effects.
\par\noindent
(10) The reference to the adopted photometry.

\smallskip\noindent
The bolometric luminosities of the surveyed areas
($L_{Bol} = 1.4 L_V$, Buzzoni 1985 private communication)
have been computed by
numerical or analytical integration of a King model (King, 1966) described by
the three observed parameters $r_c, r_t, \sigma$ --~{\it i.e.} core radius,
tidal radius and  central brightness (from Webbinnk 1985), and imposing
the boundary condition that the integral of the model brightness profile
between 0 and $r_t$ must be equal to the total cluster luminosity
$M_V$.

The adopted distances have been obtained by calibrating the observed
luminosity level of the HB, $V_{HB}$ (taken from Armandroff, 1989)
with the relation proposed by Lee \etal (1990) $M_V^{HB}=0.17[Fe/H]+0.82$,
and adopting the reddening listed by Zinn (1985).

\smallskip\noindent
{\it Important caveat:}
As already emphasized by Fusi Pecci \etal (1992), the available GC sample is
very incomplete and heterogeneous in the photometric sampling actually
achieved by the individual photometries carried out by the various
authors.

For instance, some clusters (mostly the loose ones) have
been properly surveyed down to the Main Sequence Turnoff (or even fainter)
almost uniformely. In many objects the survey has been limited to the more
external areas, and, on the contrary, in a few others the very inner regions
have been observed with HST at a much higher resolution.

This sampling bias is very insidious and must be always recalled
while discussing any result on this subject. In particular, since
strong variations in radial distributions of BSS with respect to
``normal'' stars of similar magnitude have already been repeatedly
pointed out, any comparison of BSS number as a function of
the distance from the cluster center must be done with special care,
and also the comparisons based on total numbers of detected BSS
could be strongly in error.

On the other hand, though still in its early childhood, the attempt
of looking for possible correlations of BSS with cluster properties
may be of tremendous importance to understand better the BSS and
the whole subject involving globular cluster population and
environment. Therefore, we present below and discuss the results one
can derive with the {\it available} data as a first step in a long
path which will be hopefully checked soon and corraborated by
further data. Consequently, the use of the figures listed in the tables
must be done with particular care and being aware of the possible
bias still affecting them.

\bigskip\noindent
\centerline {3. RESULTS AND DISCUSSION}

\bigskip\noindent
\centerline {\it 3.1. Playing with BSS absolute numbers}
\smallskip\noindent
In Figure 1, the number of the detected BSS --~$N_{BSS}$~-- has
been plotted as a function of the sampled light --~$L_S$, expressed in unit
of $10^{4} L_{\odot}$ ~~({\it panel a}), and versus the fraction of the
total cluster light actually sampled ~~({\it panel b}).

As can be seen from the plots, at a first inspection it is hard to find any
overall meaningful correlation and one could simply conclude that the data
are still too uncertain and scattered to draw any indication.

However, for a deeper analysis, it may be useful to divide the
cluster set into two sub-groups, based on the values of $L_S$ --~the
cluster light actually sampled in the considered observations.
So doing,  one could get some interesting hints.

\medskip\noindent
{\it (i)} $L_S < 6\times 10^4 L_{\odot}$
\par\noindent
Among the 18 GCs present in this sub-group,
$N_{BSS}$ seems to increase with $L_S$ according to the relation:
$N_{BSS} = 6.57\times L_S + 10.1$,
with a linear correlation coefficient $r=0.76$, (yielding a probability
of $\sim 99.99\%$ that the two quantities are actually correlated each other).

\medskip\noindent
{\it (ii)} $L_S > 6\times 10^4 L_{\odot}$
\par\noindent
This sub-group includes only 8 objects and its statistical significance
is therefore quite low. However, not only there is no evidence for
the existence of any trend similar to that found in the previous sub-group,
but it seems remarkable to note that all the clusters are far from
being located on the extension of the previous relation and display
a much lower number of BSS than expected if a similar incidence per
fixed sampled light is adopted.

An obvious explanation for this result could be the existence of a
bias in the BSS search which would lead to an increasing incompleteness
with increasing the sampled light. This may well be occurred, but
{\it a priori} there is no obvious reason to explain why BSS should be lost at
a larger extent while increasing the light sampling.

Alternatively, one could imagine three possible explanations: (i) the
light sampled in these clusters is not ``truly-representative'' of
the average cluster population; (ii) the light sampled in the cluster
is ``truly-representative'' of the average cluster population, but
the BSS are peculiarly distributed and the light actually sampled
is BSS-poorer than the non-sampled fraction; (iii) the clusters included
in the two considered sub-groups are ``intrinsecally'' different as far as
BSS production/survival is concerned.

To explore item (i), we refer to Fig. 1$b$. As can be seen,
the 8 clusters included in the second sub-group ({\it full dots}) display
a distribution of fractional sampled light ({\it i.e.} ${L_S}/{L_{tot}}$)
similar to those membering the first sub-group ({\it open dots}).
In particular, there are 4 objects (over 8) sampled at more than $50\%$
and only 2 at less than $20\%$. Hence, though possible, alternative (i) seems
to be improbable.

Concerning item (ii), the discussion is much more complicate. In fact,
different regions of the clusters have in general been surveyed
with different instrumental setups. Therefore, if BSS have a peculiar
distribution with respect to ``normal'' stars, one could well even
lose a whole huge BSS population if the ``wrong'' areas are sampled.

Is this the case for the 8 considered clusters? Maybe, but it is not
sure, and the answer may differ from cluster to cluster. For instance,
47 Tuc and M15 have been observed with HST in the central regions
and repeatedly observed from the ground in the outer parts. A small group
of undetected BSS may well be present, but the existence of a very
large undetected BSS population seems beyond the predictions one
could imagine on the present grounds. On the other hand, for example
NGC 2419 is a very far and rich cluster, and the MS Turnoff has been
barely observed so far. There is thus no reason to exclude that very many
BSS could have been overlooked. It would thus be curious that
``wrong'' areas or insufficient search quality have been adopted while
sampling more and more cluster light, but it cannot be excluded at
this stage.

If we assume for sake of reasoning that alternative (i) and (ii) could be
overcome, it may be eventually useful to verify whether there is at least one
parameter significantly different between the two sub-groups,
which could somehow originate an ``intrinsic'' difference between the
clusters as far as the BSS are concerned.

The most evident difference between the two sub-groups is the
cluster concentration. In fact, the mean central density is
$<Log \rho_0 = 1.5\pm0.3>$ and $<Log \rho_0 = 3.3\pm0.5>$, respectively.

Hence, though the difference is not very strong on statistical grounds,
one could have that while in loose clusters the number of the detected BSS
is proportional to the sampled light, in the intermediate-high
density globulars $N_{BSS}$ drops with respect to this relation,
and the actual number of BSS depends on various other factors
(to be found). The present data would indicate in particular
a  low number of BSS in high density clusters even
if a large fraction of the total light is sampled in the central regions.

Since it is now widely accepted (see Stryker 1993 for a review) that
BSS may originate through several processes, it is also quite
evident that, if BSS of different origin populate different clusters
(or even different regions of the same cluster, Fusi Pecci \etal 1992,
Ferraro \etal 1993), it may be natural to expect the existence of
(various) different relationships between the overall cluster properties
and the BSS frequency.

On the other hand, it is also immediate to note that, independent of
any peculiarity in the BSS radial distribution, the BSS detection
is more difficult and less ``safe'' when dealing with more and more
concentrated clusters. And thus, by explaining the detected effect as due
to cluster concentration, one could indirectly fall back to the
selection bias.

In summary, there are serious indications that the available samples
of BSS is still inadequate to allow any firm conclusions. However,
the emerging indications, if confirmed, could also be compatible with
an interpretative scenario where in loose clusters the BSS form
from primordial binaries and their number is roughly proportional to
the sampled luminosity. With increasing the cluster concentration and
the number of cluster members, primordial binaries become less and
less efficient as BSS progenitors while collisional binaries take
place, and the BSS number actually depends on a variety of parameters related
to the type of star interactions and mechanism originating the BSS
themselves.

As a consequence, it may be worth of the effort, on the one hand,
the extension of this analysis to make more direct the comparisons among
different clusters, and on the other, a deeper discussion of the
possible influence of crowding effects on the whole photometric
treatment.
To this aim, relative and normalized quantities are much better than absolute
numbers. Therefore, we define and use below a new quantity to
make more evident the bulk of the present result. Then, we deal
with crowding impact.

\bigskip\noindent
\centerline {\it 3.2. Playing with BSS normalized quantities}
\smallskip\noindent
Bolte \etal (1993) defined the BSS {\it specific frequency}
{}~~$F_{BSS}$ as the number of BSS with respect to the number of
all the observed stars brighter than two magnitudes below the
Horizontal Branch level. Though very useful in principle, this relative
quantity can hardly be precisely computed at this time in the GCs
where BSS have been detected so far since most of the photometric
surveys are not sufficiently reliable in this respect.
In fact, quite often bright stars in the observed fields are saturated
due to the long exposure needed to reach the TO region with an appropriate
S/N ratio and the completeness checks for both bright and faint stars
are insufficient or inexistent at all.

For these reasons we defined (Ferraro \etal 1993) a different
BSS {\it specific frequency} in terms of the sampled luminosity as
$ S4_{BSS} = N_{BSS}/L_S$
where $N_{BSS}$ is the number of the detected BSS and $L_S$ is the integrated
bolometric luminosity of the surveyed area in unity of $10^4 L_{\odot}$.

Admittedly, if available, the specific frequency defined by Bolte \etal
is better as it is based on ``true'' numbers and not on integrated properties.
However, our quantity has the advantage of being measurable directly on
the used frames (by properly calibrating the integrated flux collected by the
receiver) or, in absence of the necessary observational data, it
can also be (roughly) computed by knowing the cluster brightness profile
and the exact location and size of the surveyed area.

In Figure 2 we present the distribution of the $S4_{BSS}-$values obtained
for the 26 GCs listed in our catalog. In summary:
\smallskip\noindent
\item {a)} The specific frequency (as computed with the available samples)
can vary from cluster to cluster by a factor $\sim 30$.
\smallskip\noindent
\item {b)} The quite strong incidence of clusters with small specific
frequencies runs parallel to the effects due to detection losses and
selection bias. It seems quite improbable that the observed large
variation can be totally ascribed to just systematic effects, but
frankly it cannot be firmly excluded at this stage.
\smallskip\noindent
\item {c)} Taken at face value, the $S4_{BSS}$ weighted mean turns to be
$<S4_{BSS}>=10\pm 2$, where the weights have been computed taking into
account only the Poisson error in the star counts, ({\it i.e.} assuming
the uncertainty in the sampled light to be trascurable). The mean standard
deviation ($\sigma$) is quite high ($\sigma \sim 8$), and the  observed
distribution is still compatible with a Gaussian distribution.
Under this assumption,  the probability of finding 8 objects in the beam
$0 < L_S < 4$ (see Figure 2) is only $0.7\%$. Of course, this may also be taken
as another indication that a bias is surely present in the available sample.

\medskip\noindent
Assuming that the $S4_{BSS}-$values are sufficiently reliable, it is
useful to plot them versus some intrinsic structural parameters
of the BSS parent cluster.

In Figure 3{\it a,b,c} the specific frequency $S4_{BSS}$ of each cluster
has been plotted versus the integrated cluster magnitude $M_V$,
the central density $Log\rho_0$, and the concentration parameter $c$,
respectively. As can be seen, in the diagrams there are always two well
separated regions: a {\it permitted zone}, sparcely populated, and
a sort of {\it forbidden zone} completely empty. Note that an arbitrary
{\it dash line} has been reported in the plots in order to put into better
evidence the effect.

The interpretation of these plots may be risky at this stage, but also
interesting to note for future studies. In synthesis, one could see
a quite clear indication that there is a lack of clusters having
high density and/or large number of stars (bright $M_V$) and large
BSS specific frequency, while with decreasing total luminosity and/or
concentration, clusters display a quite wide set of possible $S4_{BSS}-$values.

How confident can one be that the observed effect has something to do
with reality and is not just the perverse combination of various
factors, even in presence of non-misleading data?

The problem is quite tricky since:
\smallskip\noindent
\item{i)} As shown for example in the catalog compiled by Djorgovski
and Meylan (1993), concentration and central density are well correlated
with integrated luminosity (see their Figure 2) in the sense that luminous
clusters generally have smaller cores and higher concentration. Hence, all
quantities depending on integrated luminosity turn to be automatically
correlated also with the structural parameters of the clusters.
\par\noindent
\item{(ii)} The absolute luminosity actually sampled in the considered clusters
is strongly correlated (r=0.81) with the total integrated luminosity of each
cluster (see Figure 4) as one cannot sample more light than available
in low luminosity clusters.
\par\noindent
\item{iii)} $S4_{BSS}$ is {\it by definition} strongly dependent on the
sampled luminosity and, in turn, via points (i)-(ii) on the structural
parameters. On the other hand, it is important to stress that any
specific frequency (including the one adopted by Bolte \etal, based on star
counts, see below) has this implicit dependence when plotted versus the cluster
structural parameters.

\smallskip\noindent
To explore a bit further the issue, we report in Figure 5$a$
{}~~$S4_{BSS}$ versus $M_V$ to be compared with the data in Figure 5$b$,
where the quantity $1/L_s$ referred to each cluster is plotted versus
the same abscissa.

It is quite evident that the observed number of BSS introduces only a sort
of {\it second order} perturbation to the trend driven by $L_S$.
Moreover, this depends only marginally from the sampling level of the cluster.
In fact, assuming for sake of simplicity the best observative hypothesis
({\it i.e.} the total cluster light $L_T$ actually sampled), one gets the
solid line in Figure 5$b$ ~~({\it i.e.} $1/L_T$, vs $M_V$).
In other words, since low density clusters are in general intrinsically poor,
$S4_{BSS}$ grows up easily for them even with a few detected BSS.
On the contrary, only a huge number of BSS detections could move a cluster
like 47 Tuc (highly concentrated and rich) from its position in this diagram.

On the other hand, since at present the number of BSS actually detected in
47 Tuc is small (though the cluster has been widely observed both from the
ground and with HST), the relative paucity of BSS in 47 Tuc with respect to the
typical loose clusters seems to be noteworthy.

The use of star counts to compute the specific frequency as
done by Bolte \etal (1993) leads similarly to a quantity which is
implicitly related to the cluster luminosity and stellar density
as the numbers of giant branch and horizontal branch stars detectable
in a given region of the cluster are proportional to the cluster sampled
light (see the so-called ``Fuel Consumption Theorem'',
Renzini and Buzzoni 1986).

To directly verify this and to compare the specific frequency we
defined with the one adopted by Bolte \etal 1993, we have computed and listed
in Table 2 the specific frequencies computed for 5 clusters where all the
needed quantities are available to us in a computer readable form.
In addition, we have also listed in column 4 the number of HB plus
RGB stars (brighter than the HB level) observed
and those predicted by inserting our estimated sampled luminosity
in the relation (Renzini and Buzzoni, 1986):
$$N_j = B(t) L_S t_j$$
where $L_S$ is the sampled luminosity (see column 6 of Table 1),
$t_j$ is the duration
of the considered phase, and $B(t)$ is the evolutive flux.
In the computations we have assumed from Renzini and Buzzoni (1986) and
Buzzoni \etal (1983) $B(t)=2\times 10^{-11}$, $t_{HB}=10^8 ~~yr$,
and $t_{RGB}=7.3\times 10^7 ~~yr$ (corresponding to the RGB lifetime
spent above the HB level).

As can be seen from the Table 2, the numbers of predicted and observed
stars are in excellent agreement, confirming that our adopted
procedure to estimate $L_S$ is reliable enough to avoid any bias. Moreover,
the plot presented in Figure 6, reporting our specific frequency
--~$S4_{BSS}$~-- versus the similar quantity determined by using
the definition of Bolte \etal (1993), shows that the two observables
are strictly correlated, as expected. Hence, the use of the BSS specific
frequencies to study the possible existence of clear-cut correlations
between them and the parent cluster overall properties is really
informative only if they are related to parameters which are fully
independent of the cluster absolute magnitude.

\bigskip\noindent
\centerline {\it 3.3. Crowding effects: further caveats}
\smallskip\noindent
Since it is immediate to immagine that the available data must be influenced
by crowding, it is useful to add some information and a greater discussion
of its possible effects on our procedure.
\par
In the search for BSS candidates there are at least three basic aspects
which may introduce significant biases in the samples:
\smallskip\noindent
\item{1.} The quality of the seeing or, more in general, the actual
spatial resolving power during the observations.
\par\noindent
\item{2.} The magnitude limit actually reached in the photometry, {\it
i.e.} how faint and precise are the measures down to the cluster turnoff.
\par\noindent
\item{3.} The intrinsic crowding in the cluster sampled region, which
is function of the cluster structural parameters and, for given
observational setup, of the cluster distance.
\medskip\noindent
Concerning the seeing, it is quite obvious that it is really discriminant
only when the properties mentioned in items 2 and 3 are the same.
We collected the very poor information on the seeing conditions of the
CMD's used in Table 1 and found that they range from 0.5 up to 2 arcseconds.
Curiously enough, a plot of the quoted seeing (usually the FWHM of
the average star of the average frame) versus the number of detected
BSS displays in case an anti-correlation (many BSS detected in clusters
observed in bad seeing conditions!). This result shows that the detection
of BSS candidates has been so far only marginally dependent on the
actual seeing conditions (probably because intermediate-quality nights
have been generally used to study loose clusters), and suggest that
the most important factors have actually been the deepness and the
degree of completeness achieved in the photometric surveys.
\par
In this respect, it may be interesting to recall the case of the cluster
IC 4499. In a forthcoming study (Ferraro \etal 1994), we observed the
same cluster region as observed by Sarajedini (1993), in very similar
seeing conditions {\it but} we reached about two magnitudes fainter and
found almost twice as many BSS candidates as Sarajedini.
\par
To study further the possible impact of crowding, we have computed the
{\it crowding parameter} ~~$F$~~ defined as:

$$F = {A_{s_i}}/{A_{frame}}$$

\noindent
with $A_{s_i} = 80 \times L_S \times \pi \times FWHM^{2} ~~(arcsec^2)$,
$A_{frame} =$ observed area in $arcsec^2$, and where 80 is the
conversion coefficient (star/luminosity) computed via the quoted
Fuel-Consumption Theorem (Renzini and Buzzoni 1986), $L_S$ is the
sampled luminosity as determined above, and $FWHM$ is the adopted
figure for the seeing conditions in the adopted cluster photometry.
\par
In synthesis, since $F$ yields an estimate of the fraction of the frame
actually covered by stars, it should roughly rank the clusters listed in
our sample according to both intrinsic crowding and seeing impact.
\par
Figure 7 reports the distributions of the absolute ($N_{BSS}$) and
specific ($S4_{BSS}$) numbers of BSS candidates with varying
the corresponding available $F$ values.
\par
At first sight, inspection of the plots leads to two main conclusions:
\smallskip\noindent
\item{a)} The number of BSS candidates actually detected so far in the
clusters is sufficiently uncorrelated to $F$ {\it i.e.} $N_{BSS}$ do not
show any clear-cut trend with varying $F$.
\par\noindent
\item{b)} The {\it specific frequency} $S4_{BSS}$ seems to be correlated
to $F$, in the sense that the specific frequencies increase with improving
the crowding conditions.
\smallskip\noindent
This evidence could reinforce the idea that $S4_{BSS}$ is the critical
parameter to determine since we expect that the results of the BSS search
must be influenced by crowding and seeing. On the other hand, we cannot
conclude based on this plot that there is surely a bias induced by crowding
 in the available
data because of the quoted implicit dependence on $L_S$ of both $S4_{BSS}$
and $F$.
\par
In fact, $F\propto L_S$ and $S4_{BSS} \propto 1/{L_S}$. Hence, if $L_S
\rightarrow 0$ just a few BSS automatically lead towards ``high''
$S4_{BSS}$ values, with very small figures for $F$.

\bigskip\noindent
\centerline {4. CONCLUSIONS AND FUTURE PROSPECTS}

\bigskip\noindent
Using the available data on BSS in Galactic GCs and being fully aware
of the possible existence of important incompleteness and bias in
the adopted sample, we have here started a comparison between the
various parent clusters based on absolute BSS numbers and corresponding
normalized quantitities.

The number of BSS detected so far seems to increase almost linearly
with increasing the amount of sampled light in loose clusters, while
the trend changes abruptly for clusters having intermediate-high
central densities. In particular, highly concentrated objects have
much less BSS detected so far per unit luminosity than loose
globulars. However, this evidence could also be ascribed to the
increasing difficulty to detect BSS candidates with increasing
cluster concentration.

If confirmed by further data, this observational result is very
important to understand better the BSS origin and the whole
complicate interplay taking place within each cluster between
binary formation, evolution, and survival and the comprehensive
dynamical evolution of the global cluster environment.

In particular, an interpretative framework which explains the
BSS in loose clusters as produced from primordial binaries and
those detected in high density globulars as due to star
interactions leading to binary formation, merging. etc.
seems compatible with these early results.

A systematic search for BSS, spanning from the very central regions
of the clusters to their outskirts, is a crucial prerequisite
to make a significant breakthrough on this interesting issue.

\bigskip
\bigskip

\noindent{\sl Acknowledgements.}~~
FRF acknowlwdges the Visitor Program at the STScI for hospitality during
the period in which part of this work was carried out.

\bigskip
\bigskip
\parskip=0pt

\centerline{\bf References}     %%  no more than 30 for Letters to Nature !!
\bigskip

\pp Anthony-Twarog, B.J., Twarog, B.A., 1992, AJ 103, 1264.

\pp Armandroff, T.E., 1989, AJ,  97, 375.

\pp Auri\`ere, M., Ortolani, S.,
and Lauzeral, C., 1990, Nature, 344, 638.

\pp Bailyn, C.D., 1993,  in Dynamics of Globular Clusters, eds S. Djorgovski
and G. Meylan, ASP Conference series 50, p. 191,

\pp Bailyn, C.D., Grindlay, J.E., Cohn, H.N., Lugger, P.M., Stetson, P.,
 Hesser, J.E., 1989,  AJ  98, 882.

\pp Bailyn, C.D., Sarajedini, A., Cohn, H., Lugger, P.M., Grindlay, J.E., 1992,
 AJ, 103, 1564.

\pp Bailyn, C.D., Mader, V., 1993, in ``Blue Stragglers'',
ed. M. Livio and R.  Saffer, ASP conf. Series 53, p. 177.

\pp Bolte, M, 1992, ApJS 82, 145.

\pp Bolte, M, Hesser, J.E., Stetson, P.B., 1993, ApJ 408, L89.

\pp Brewer, J.P., Fahlmann, G.G., Richer, H.B., Searle, L.,
Thompson, I., 1993, AJ 105, 2158.

\pp Buonanno, R., Corsi, C.E.,  Fusi Pecci, F., 1985, A\&A 145, 97.

\pp Buonanno, R., Caloi, V., Castellani, V., Corsi, C.E., Fusi Pecci, F.
   and Gratton, R.G. 1986, A\&AS 66, 79.

\pp Buonanno, R., Buscema, G., Fusi Pecci, F., Richer, H.B., and Fahlman,
G.G., 1990, AJ 100, 1811.

\pp Buonanno, R., Fusi Pecci, F., Cappellaro, E., Ortolani, S., Ritchler, R.T.,
Geyer, H.E. 1991, AJ, 102, 1005.

\pp Buonanno, R., Corsi, C.E., Buzzoni, A., Cacciari C., Ferraro, F.R.
Fusi Pecci, F., 1994, A\&A (in press).

\pp Buzzoni, A., Fusi Pecci, F., Buonanno, R.,
Corsi, C., 1983, A\&A 123, 94.

\pp Carney, B.W., Kellar Fullton, L., Trammell, S.R., 1991, AJ 101, 1699.

\pp Cederbloom, S., Moss, M., Cohn, H., Lugger, P., Bailyn, C., Grindlay, J.,
    McClure, R. 1992, AJ, 103, 480.

\pp Christian, C.A., and Heasley, J.N., 1986, AJ  303, 216.

\pp Christian, C.A., and Heasley, J.N., 1988, AJ 95, 1422.

\pp Christian, C.A., Heasley, J.N., 1991, AJ 101, 967.

\pp Cot\'e, P., Richer, H.B., and Fahlman, G.G., 1991, ApJ 102, 1358.

\pp De Marchi, G., Paresce, F., Ferraro, F.R., 1993, ApJS 85, 293,

\pp Djorgovski, S.G., Piotto, G., King I.R., 1988, in  Dynamics of Dense
Stellar
System, ed. D. Merritt, (Cambridge Univ. Press), p. 147.

\pp Djorgovski, S.G., Meylan, G., 1993, (preprint).

\pp Ferraro, F.R., Clementini, G., Fusi Pecci, F., Buonanno, R., 1991,
MNRAS 252, 357.

\pp Ferraro, F.R., Clementini, G., Fusi Pecci, F., Sortino, R.,
 and Buonanno, R. 1992a, MNRAS, 256, 391,

\pp Ferraro, F.R., Fusi Pecci, F. Buonanno, R., 1992b,
MNRAS 256, 376.

\pp Ferraro, F.R., Paresce, F., 1993, AJ, 106, 154.

\pp Ferraro, F.R., Fusi Pecci, F., Cacciari, C., Corsi, C.E., Buonanno, R.,
Fahlman, G.G., Richer, H.B., 1993, AJ 106, 2324.

\pp Ferraro, F.R., {\it et al}, 1994. (to be submitted).

\pp Fusi Pecci, F., Ferraro, F.R., Corsi, C.E., Cacciari, C., and
Buonanno, R. 1992, AJ, 104, 1831.

\pp Fusi Pecci, F., Ferraro, F.R., Bellazzini, M. Djorgovski, S.G.,
Piotto, G., and Buonanno, R. 1993a, AJ, 105, 1145.

\pp Fusi Pecci, F., Ferraro, F.R., Cacciari, C., 1993b, in ``Blue Stragglers'',
ed. M. Livio and R.  Saffer, ASP conf. Series 53, p. 97.

\pp Gratton, R.G., and Ortolani, S., 1984, A\&ASS 57, 177.

\pp Guhathakurta, P., Yanny, B., Schneider, D.P., Bahcall, J.N., 1992,
AJ 104, 1790.

\pp Harris, H.C., 1993, AJ 106, 604.

\pp Hodder, P.J.C., Nemec, J.M., Richer, H.B. Fahlman, G.G., 1992,
 AJ 103, 460.

\pp Holland, S., and Harris, W.E., J.N., 1992, AJ 103, 131

\pp Kaluzny, J., Krzemiski, W., 1993, Acta Astronomica, in press.

\pp  King, I. R. 1966, AJ, 71, 64

\pp Lauzeral, C., Ortolani, S., Auri\`ere, M., Melnick, J.1993, A\&A 262, 63.

\pp Lee, Y.W., Demarque, P., Zinn, R., 1990, AJ 350, 155.

\pp Leonard, P.J.T., 1989, AJ 98, 217.

\pp Nemec, J.M., 1989, In  ``The use of Pulsating Stars in Fundamental
Problems  in Astronomy", IAU Colloquium No. 111, eds. E.J. Schmidt
(Cambridge Univeristy Press), p. 215.

\pp Nemec, J.M. 1991, Nature, 352, 286

\pp Nemec, J.M., and Cohen, J.G., 1989, ApJ 336, 780.

\pp Nemec, J.M., and Harris, H.C., 1987, AJ 316, 172.

\pp Piotto, G., King I.R., Djorgovski, S.G., 1988, AJ, 96, 1918

\pp Paresce, F.,Shara, M., Meylan, G., Baxter, D., Greenfield, P.,
Jedrzejewski, R., Nota, A., Sparks, W.B., Albrecht, R., Barbieri, C.,
Blades, J.C., Boksenberg, A., Crane, P., Deharveng, J.M., Disney,
M.J., Jakobsen, P., Kamperman, T.M., King, I.R., Macchetto, F.,
Mackay, C.D., and Weigelt, G. 1991, Nature, 352, 297

\pp Renzini, A., Buzzoni, A., 1986, In Spectral Evolution of Galaxies,
eds. C. Chiosi, A. Renzini, (Dordrecht: Reidel), p.135.

\pp Sandage, A. 1953. AJ 58, 61.

\pp Sarajedini, A.,and Da Costa, G.S., 1991, AJ 102, 628.

\pp Sarajedini, A.,  1993, AJ 105, 2172.

\pp Seitzer, P., Carney, B.W., 1990, AJ 99, 229.

\pp Smith, G.H., McClure, R.D., Stetson, P.B., and Hesser, J.E., 1986,
AJ 91, 842.

\pp Stetson, P.B., VandenBerg, D.A., Bolte, M., Hesser, J.E.,
Smith, G.H., 1989, AJ 97, 1360.

\pp Stetson, P.B., 1991, In Precision Photometry, eds. A.G.D. Philip, A. Upgren
and K.A. Janes (Davis, Schenectady), p. 69.

\pp Stryker, L.L., 1993,PASP, in press.

\pp Webbink, R.F., 1985, {\it Dynamics of Star Clusters}: ed. J. Goodman
and P. Hut (Dordrecht: Reidel), p. 541.

\pp Zinn, R.J., 1985,  ApJ 293, 424.

\parskip=5pt

\vfill\eject

%%%%%  not more than 4 figures for Letters to Nature!!

\centerline{\bf Figure Captions:}
\bigskip

%%%%% in Fig 1, comment on the scaling for the three panels, depending on
%%%%% whether the grayscale or the surface plots are used.

\bigskip
\noindent{\bf Figure 1.}~
The observed number of Blue Straggler Stars ($N_{BSS}$) as a function of the
sampled light ($L_S$): expressed in  unity of $10^4 L\odot$ ({\it panel a)}
and as percentage of the total cluster light ({\it panel b)}.
The position of two clusters (NGC 5272 and NGC 2419) are indicated
by arrows since their coordinates are outside the axis range.
The vertical dashed line in {\it panel (a)}
divides the two sub-samples showing  different behaviour with
increasing $L_S$ (see text). The {\it full dots} in {\it panel (b)}
mark clusters with $L_S > 6\times 10^4 L_{\odot}$.

\bigskip
\noindent{\bf Figure 2:}~
Histogram of the distribution of $S4_{BSS}$ in the Galactic GCs listed
in our catalog.

\bigskip
\noindent{\bf Figure 3:}~$S4_{BSS}$ vs cluster structural parameters:
{\it panel (a)} the integrated absolute magnitude, $M_V$;
{\it panel (b)} the logarithm of the central stellar density, $Log \rho_0$;
{\it panel (c)} the concentration parameter $c$.
The {\it dashed line} has been arbitrarily traced to put into evidence
the effect discussed in the text.

\bigskip
\noindent{\bf Figure 4.}~Plot of the cluster total light ($L_T$) versus the
sampled light, $L_S$. Note the tigh correlation (r=0.81) existing between the
two quantities. The object indicated by an arrow is NGC 5272 (whose actual
coordinates: 40,50).

\bigskip
\noindent{\bf Figure 5:}~
The bias affecting different BSS specific quantities as function of the
absolute magnitude $M_V$:
\par\noindent
{\it panel (a)} - The $S4_{BSS}$ parameter.
\par\noindent
{\it panel (b)} - The inverse of the sampled light ($1/L_S$). As can be seen,
the trend showed in {\it panel (a)} is partially due to the relationship
necessarily linking $L_S$ and $M_V$. The {\it solid line} represent
the inverse of the total cluster light.

\bigskip
\noindent{\bf Figure 6:}~
Comparison between the  BSS specific frequency defined in this paper
($S4_{BSS}$) and $F_{BSS}$ (Bolte \etal 1993).

\bigskip\noindent

{\bf Figure 7:}~
The crowding parameter $F$ opposed to $N_{BSS}$ ({\it lower panel}) and to
$S4_{BSS}$ ({\it upper panel}). Note that $F$ increase as the crowding
condition
worsen (see text).
\bye

%%%%%tab1

\input basicnew
\newdimen\digitwidth
\setbox0=\hbox{\rm0}
\digitwidth=\wd0

\vsize=29truecm
\hsize=20truecm
\nopagenumbers
\voffset=-6truemm
\tabskip=2em plus1em minus1em
{\bf Table 1.} Adopted parameters for globular clusters
containing Blue Stragglers.
\trule
\halign to
\hsize{#\hfil&\hfil#\hfil&\hfil#&\hfil#&\hfil#\hfil&\hfil#\hfil
&\hfil#\hfil
&\hfil#\hfil
&\hfil#\hfil
&\hfil#\hfil\cr
 $Name$ &$ Log(\rho_0)$ & $c$ & $M_V$  & $[Fe/H]$
&~$L_{S}$ & $N_{BSS}$ & $S4_{BSS}$ &$F$ &
$Reference$ \cr
\mrule
Pal 15~~ &-0.62&0.60& -5.20&-1.90&~0.9&~10&11.24 &0.007&Seitzer \& Carney (1990
) \cr
Pal 5~~  &-0.49&0.80& -5.03&-1.47&~0.2&~~8&33.33 &--&Smith {\it et al.} (1986)
\cr
Pal 14~~ &-0.46&0.79& -4.83&-1.47&~0.4&~10&25.00 &--&Holland \& Harris (1992)
\cr
Pal 4~~~ &~0.08&0.76& -6.11&-2.20&~2.3&~12&~5.33 &0.025&Christian \& Heasley
(1986) \cr
Pal 3~~~ &~0.27&0.96& -6.11&-1.78&~2.4&~~9&~3.67 &--&Gratton \& Ortolani (1984)
 \cr
NGC 5053 &~0.58&0.75& -6.31&-2.58&~2.6&~24&~9.23 &0.005&Nemec \& Cohen (1989)
\cr
NGC 7492 &~1.06&0.97& -5.64&-1.51&~1.1&~27&25.47 &0.005&Cot\`e {\it et al.} (19
91) \cr
Rup 106~ &~1.22&1.08& -6.22&-1.09&~2.1&~32&15.24 &0.004&Buonanno {\it et al.}
(1990) \cr
NGC 5466 &~1.25&1.12& -7.23&-2.22&~5.8&~48&~8.28 &--&Nemec \& Harris (1987)
\cr
Pal 12~~ &~1.38&1.35& -4.78&-1.14&~0.8&~21&26.92 &0.001&Stetson {\it et al.}
(1989)\cr
NGC 2419 &~1.52&1.41& -9.70&-2.10&49.0&~11&~0.22 &--&Christian \& Heasley
(1988)  \cr
IC 4499  &~1.63&1.10& -7.70&-1.50&~7.9&~24&~3.04&0.032&Sarajedini (1993)\cr
NGC 5897 &~1.81&0.95& -6.19&-1.75&~1.3&~34&10.30 &0.039&Ferraro {\it et al.}
(1992b)\cr
NGC 288~ &~2.02&0.97& -6.71&-1.40&~2.9&~38&13.10 &0.011&Bolte (1992)\cr
NGC 4372 &~2.33&1.08& -7.74&-2.08&~7.4&~35&~4.73 &0.015&Kaluzny \& Krzeminski
(1992)  \cr
NGC 6366 &~2.40&0.96& -6.12&-0.71&~1.2&~27&22.50 &--&Harris (1993)  \cr
NGC 6101 &~2.80&1.30& -6.93&-1.80&~2.2&~27&12.27 &0.008&Sarajedini \& Da Costa
(1991)\cr
NGC 3201 &~3.12&1.52& -7.57&-1.56&~2.3&~31&12.92 &--&Breweer {\it et al.}
(1993)\cr
NGC 5024 &~3.17&1.73& -8.99&-2.04&~7.0&~~4&~0.57 &--&Christian \& Heasley
(1991)  \cr
NGC 6838 &~3.24&1.22& -5.68&-0.58&~0.5&~18&19.57 &0.002&Hodder {\it et al.}
(1992)\cr
NGC 6171 &~3.27&1.50& -6.88&-0.89&~1.9&~26&11.05 &0.003&Ferraro {\it et al.}
(1991)\cr
NGC 6229 &~3.40&1.51& -8.29&-1.54&21.3&~13&~1.62 &--&Carney {\it et al.} (1991)
\cr
NGC 5272 &~3.85&1.97& -9.19&-1.66&40.0&137&~3.42 &--&B94 + FF94 + S53\cr
NGC 6397 &~4.20&~1.70& -6.75&-1.91&~2.0&~24&12.00 &--&A90 + ATW92\cr
NGC 104~ &~5.02&2.06& -9.53&-0.71&11.0&~24&~2.18 &--&Guhathakurta {\it et al.}
(1992)\cr
NGC 7078 &~5.26&2.36& -9.39&-2.15&14.0&~~9&~0.64 &--&Ferraro \& Paresce (1993)
\cr }\trule
\par\noindent
B94 $=$ Buonanno {\it et al.} (1994)
\par\noindent
FF94 $=$ Ferraro {\it et al.} (1994)
\par\noindent
S53 $=$ Sandage (1953)
\par\noindent
A90 $=$ Aurier\`e {\it et al.} (1990)
\par\noindent
ATW92 $=$ Anthony-Twarog \&Twarog (1992)
\par\noindent

\par\noindent
\par\noindent
\par\noindent
\medskip
\vfill\eject
\bye

%%%%%%%%%tab2

\input basicnew
\newdimen\digitwidth
\setbox0=\hbox{\rm0}
\digitwidth=\wd0

\vsize=29truecm
\hsize=16truecm
\nopagenumbers
\voffset=-6truemm
\tabskip=2em plus1em minus1em
{\bf Table 2.} BSS frequencies and observed and expected
populations on RGB and HB.
\trule
\halign to
\hsize{#\hfil&\hfil#\hfil&\hfil#\hfil
&\hfil#\hfil
&\hfil#\hfil&#\hfil\cr
 $Name$ &
$S4_{BSS}$&
$F_{BSS}$ &
$(N_{RGB}+N_{HB})_{obs}$ & $
(N_{RGB}+N_{HB})_{exp}$  & $Reference$ \cr
\mrule
NGC 7492 &25.5~& -- & ~45 &~38 & Cot\`e {\it et al.} (1991) \cr
Rup 106~ &15.3~&0.17 & ~88 &~73 & Buonanno {\it et al.} (1990) \cr
NGC 5272 &~~2.0$^{\dag}$ &0.02 & 390 &~400 & Buonanno {\it et al.} (1994) \cr
NGC 5897 &10.3~&0.11 & 113 & 115&Ferraro {\it et al.} (1992b)\cr
NGC 6171 &13.7~&0.11 & ~74  & ~66 &Ferraro {\it et al.} (1991)\cr
IC 4499 &~3.0~&0.05 & --  & -- &Sarajedini  (1993)\cr
}\trule
$^{\dag}$ Computed only on the Bright Photographic Sample (Buonanno
{\it et al.} 1994)
\par\noindent
\par\noindent
\par\noindent
\medskip
\vfill\eject
\bye